\documentclass{article}
\usepackage{waspaa21,times}
\usepackage{url}

\usepackage{amsmath,graphicx, color}
\usepackage{epstopdf, latexsym, amssymb, cite, ushort, mathtools, caption, url, pgfplots, tikzscale}
\usepackage{booktabs}
\usepackage{ushort}
\usepackage{calrsfs}
\DeclareMathAlphabet{\pazocal}{OMS}{zplm}{m}{n}
\usepackage{makecell}
\usepackage{array}
\usepackage{textcomp}
\usepackage{calc}
\usepackage{tikz}
\usetikzlibrary{shapes,arrows}
\usetikzlibrary{calc}
\usetikzlibrary{positioning}
\usetikzlibrary{arrows}
\usetikzlibrary{arrows.meta}
\usetikzlibrary{arrows,scopes}
\usetikzlibrary{backgrounds}
\usetikzlibrary{patterns}
\usepackage{xfrac} % diagonal fraction
\usepackage{nicefrac}  % diagonal fraction
\usepackage{upgreek} % upright greek letters
\usepackage{cases} % for numbered cases environment
\usepackage{subcaption} % for subfigure
\usepackage{siunitx}
\usepackage{empheq}
\usepackage{xcolor}
\usepackage[font=small]{caption}

\def\BibTeX{{\rm B\kern-.05em{\sc i\kern-.025em b}\kern-.08em
    T\kern-.1667em\lower.7ex\hbox{E}\kern-.125emX}}

\DeclareMathOperator {\sinc}{sinc}

\DeclareMathOperator {\tr}{tr}

\graphicspath{{./fig/}}

\allowdisplaybreaks

\def\h{\mathbf{h}}

\def\0{{\mathbf{0}}}

\def\bPsi{\boldsymbol{\Psi}}

\newcommand*{\transp}{{\scriptscriptstyle{T}}}
\newcommand*{\herm}{{\scriptscriptstyle{H}}}

\newlength\fheight
\newlength\fwidth

\title{Low-Complexity Steered Response Power Mapping based on Nyquist-Shannon Sampling}

\name{Thomas Dietzen,$^{1}$
      Enzo De Sena,$^{2}$
      Toon van Waterschoot$^{1}$\sthanks{This research work was carried out at the ESAT Laboratory of KU Leuven, in the frame of KU Leuven internal funds C24/16/019 "Distributed Digital Signal Processing for Ad-hoc Wireless Local Area Audio Networking" and VES/19/004, and FWO Large-scale research infrastructure "The Library of Voices - Unlocking the Alamire Foundation’s Music Heritage Resources Collection through Visual and Sound Technology" (I013218N). The research leading to these results has received funding from the European Research Council under the European Union's Horizon 2020 research and innovation program / ERC Consolidator Grant: SONORA (no. 773268). This paper reflects only the authors' views and the Union is not liable for any use that may be made of the contained information.}}
      
\address{$^1$ KU Leuven, Dept. of Electrical Engineering (ESAT), STADIUS Center for Dynamical Systems,\\ Signal Processing and Data Analytics, Leuven, Belgium\\ \{thomas.dietzen, toon.vanwaterschoot\}@esat.kuleuven.be\\              
         $^2$ University
of Surrey, Institute of Sound Recording, Guildford, UK\\
          e.desena@surrey.ac.uk
}

\begin{document}

\ninept
\maketitle

\begin{sloppy}

\begin{abstract}
The steered response power (SRP) approach to acoustic source localization computes a map of the acoustic scene from the frequency-weighted output power of a beamformer steered towards a set of candidate locations.
Equivalently, SRP may be expressed in terms of time-domain generalized cross-correlations (GCCs) at lags equal to the candidate locations' time-differences of arrival (TDOAs).
Due to the dense grid of candidate locations, 
%each of which inverse Fourier transforms (IFTs) are evaluated for, 
each of which requires inverse Fourier transform (IFT) evaluations,
conventional SRP exhibits a high computational complexity. 
In this paper, we propose a low-complexity SRP approach based on Nyquist-Shannon sampling.
Noting that on the one hand the range of possible TDOAs is physically bounded, while on the other hand the GCCs are bandlimited, 
we critically sample the GCCs around their TDOA interval and approximate the SRP map by interpolation.
In usual setups, the number of sample points can be orders of magnitude less than the number of candidate locations and frequency bins, yielding a significant reduction of IFT computations at a limited interpolation cost.
Simulations comparing the proposed approximation with conventional SRP indicate low approximation errors and equal localization performance.
\mbox{MATLAB} and \mbox{Python} implementations are available online.
\end{abstract}


\section{Introduction}

Acoustic source localization \cite{chen2010introduction, dibiase2000high} in noisy and reverberant environments remains an essential task in many multi-microphone signal processing \cite{brandstein2013microphone, benesty2008microphone} applications.

A localization approach that is said to be robust to such adverse conditions is known as steered response power (SRP) mapping \cite{dibiase2000high, velasco2016proposal}.
Here, a map of the acoustic scene is computed from the frequency-weighted output power of a delay-and-sum (DSB) beamformer steered towards a set of candidate locations. 
Given this map, the location of a single source may be inferred from the maximum.
It is well known \cite{dibiase2000high, velasco2016proposal} that SRP may equivalently be expressed in terms of time-domain generalized cross-correlations (GCCs) \cite{knapp1976generalized}  of the microphone signals at lags equal to the candidate locations' time-differences of arrival (TDOAs).
%To obtain these, inverse Fourier transform (IFT) evaluations are required per candidate location and microphone pair.
The major disadvantage of conventional SRP is posed by its high computational complexity due to the dense grid of candidate locations, each of which requires inverse Fourier transform (IFT) evaluations.

In the literature, several low-complexity SRP approaches \cite{zotkin2004accelerated, dmochowski2007generalized, do07, do2007fast, do09, cobos2010modified, marti13, nunes14, lima2014volumetric, grondin2019svd} have been proposed. 
In stochastic region contraction \cite{do07, do09} or coarse-to-fine search strategies \cite{zotkin2004accelerated, do2007fast}, the search space is iteratively contracted based on the  SRP map obtained in the previous iteration.  
Volumetric approaches \cite{cobos2010modified, marti13, nunes14, lima2014volumetric} accumulate the SRP function over the GCCs corresponding to the volume
surrounding the candidate locations.
This allows for a coarser spatial grid \cite{cobos2010modified, marti13, nunes14} and (similarly to \cite{zotkin2004accelerated, do07, do09, do2007fast}) enables iterative \cite{marti13} or hierarchical \cite{nunes14} search approaches. 
In \cite{dmochowski2007generalized}, instead of searching over space, the search is performed over TDOAs, at each of which the SRP map is updated for all corresponding candidate locations. 
In \cite{grondin2019svd}, a low-rank approximation of the SRP projection matrix is proposed.

In this paper, we propose a low-complexity SRP approach based on Nyquist-Shannon sampling \cite{marks2012introduction}.
We exploit the fact that on the one hand the range of possible TDOAs is physically bounded by the inter-microphone distance and the speed of sound, while on the other hand the GCCs are bandlimited. 
This allows us to critically sample the GCCs around their TDOA interval and approximate the SRP map by interpolation.
Here, the number of auxiliary sample points outside the interval is a design parameter, which determines the quality of the approximation.
In usual setups, the number of sample points can be orders of magnitude less than the number of candidate locations and frequency bins, yielding a significant reduction of IFT computations at a limited interpolation cost.\footnote{Since the proposed approach allows to approximate the entire SRP map from GCC samples, it may potentially be combined with  refined spatial search approaches \cite{zotkin2004accelerated, do07, do2007fast, do09, cobos2010modified, marti13, nunes14} for further complexity reduction. Such extensions are, however, beyond the scope of this paper.} 
Simulations comparing the proposed approximation with conventional SRP indicate low approximation errors and equal localization performance using only few auxiliary samples.
\mbox{MATLAB} and \mbox{Python} implementations are available online \cite{Waspaa21Code, Waspaa21CodePython}.
Note that while GCC interpolation based on Nyquist-Shannon sampling \cite{qin2008subsample} and other methods \cite{tervo2008interpolation} has been studied in the past, it has so far not been applied in the context of SRP complexity reduction.

We revisit the conventional SRP approach in Sec.~\ref{sec:SRP}.
In Sec.~\ref{sec:compred}, we propose the low-complexity SRP approach based on Nyquist-Shannon sampling.
In Sec.~\ref{sec:sim}, the proposed approach is evaluated against the conventional approach.
Sec.~\ref{sec:conclusion} concludes the paper.   
 
\section{Steered Response Power Mapping}
\label{sec:SRP}

The SRP map is defined over a set of candidate locations.
Towards each candidate location, one steers a DSB and obtains the SRP value as the frequency-weighted output power. 
We define the signal and propagation model in Sec. \ref{sec:sigpropmodel}, revisit SRP in Sec.  \ref{sec:bfperspective}, outline its relation to the time-domain GCC in Sec. \ref{sec:timedomainGCC}, and discuss its complexity in Sec. \ref{sec:compcomp}.

Notation: vectors and matrices are denoted by lower- and upper-case boldface letters, $\mathbf{A}^*$, $\mathbf{A}^\transp$, $\mathbf{A}^\herm$, $\Re[\mathbf{A}]$, $\tr[\mathbf{A}]$, and $[\mathbf{A}]_{m,m'}$ respectively denote the complex conjugate, the transpose, the Hermitian, the real part, the trace, and the element of $\mathbf{A}$ at index $(m,m')$.
The imaginary unit is denoted by $j$.
\subsection{Signal and Propagation Model}
\label{sec:sigpropmodel}
In the frequency domain, with $\omega$ the radial frequency and $M$ the number of microphones, let $y_m(\omega)$ with $m = 1,\dots M$ denote the $m^{\text{th}}$  microphone signal, which is assumed to be bandlimited by $\omega_0$.
In SRP, the DSB is usually not directly applied to the microphone signals, but to a weighted version thereof in order to improve spatial resolution \cite{dibiase2000high, knapp1976generalized, velasco2016proposal, zhang2008does}. 
To this end, one defines the deterministic frequency-domain GCC matrix $\bPsi_{{y}}(\omega) \in \mathbb{C}^{M\times M}$ as
\begin{align}
[\bPsi(\omega)]_{m,m'} &= \psi_{m,m'}(\omega)\nonumber\\
&=\gamma_{m,m'}(\omega)y_m(\omega)y_{m'}^*(\omega), \label{corrmatrix}
%\boldsymbol{\ell}_{\textsl{m}|1} - \boldsymbol{\ell}_{\textsl{s}|i}
\end{align}
where $ \gamma_{m,m'}(\omega)$ denotes the weighting function.
A common, heuristically motivated weighting approach is the so-called phase transform (PHAT) \cite{dibiase2000high, knapp1976generalized, velasco2016proposal, zhang2008does, zotkin2004accelerated, dmochowski2007generalized, do07, do2007fast, do09, cobos2010modified, marti13, nunes14} given by $\gamma_{m,m'}(\omega) = {\displaystyle 1}/{\displaystyle \lvert y_m(\omega)y_{m'}^*(\omega)\rvert}$, in which case only phase information remains on the right-hand side of (\ref{corrmatrix}). 
The remainder of this paper holds regardless of the actual weighting approach.
 
Let the Cartesian coordinate vectors of the $M$ microphone locations and the $J$ candidate locations be denoted as $\mathbf{p}_{m}$ and  $\mathbf{q}_{i}$ with $i = 1, \dots,\, J$, respectively.
The TDOA $\Delta t_{m,m'}(i)$ of a sound wave originating from $\mathbf{q}_{i}$ as observed at the microphone pair $(m,m')$ is given by
\begin{align}
 \Delta t_{m,m'}(i) = (\lVert \mathbf{p}_{m} - \mathbf{q}_{i} \rVert - \lVert  \mathbf{p}_{m'} - \mathbf{q}_{i} \rVert)/c, \label{eq:Deltat_i}
\end{align}
where $c$ is the speed of sound.
If far field propagation is assumed,
 the candidate location may be expressed by the incident direction vector of unit length, in the following denoted by $\mathbf{r}_{i}$. 
In this case, the TDOA $\Delta t_{m,m'}(i)$ becomes
\begin{align}
\Delta t_{m,m'}(i) &= (\mathbf{p}_{m} -  \mathbf{p}_{m'})^\transp\mathbf{r}_{i}/c. \label{eq:Deltat_i_spher}
\end{align}
Note that with  $d_{m,m'} = \lVert  \mathbf{p}_{m} - \mathbf{p}_{m'} \rVert$ the inter-microphone distance, we find that 
\begin{align}
\Delta t_{m,m'}(i)   &\in [-\Delta t_{m,m'|0}  ,\, \Delta t_{m,m'|0} ] \label{eq:Deltat_i_range} \\
\text{with}\quad\Delta t_{m,m'|0} &=d_{m,m'} /c,
\end{align}
where the limits of the interval are obtained if $\mathbf{p}_{m}$, $\mathbf{p}_{m'}$, and  $\mathbf{q}_i$ lie on the same line or if  $\mathbf{p}_{m} - \mathbf{p}_{m'}$ and $\mathbf{r}_{i}$ are codirectional. 
The number of different microphone pairs $(m,m')$ is given by
\begin{align}
P = M(M-1)/2.
\end{align}

\subsection{Steered Response Power}
\label{sec:bfperspective}

The DSB steering vector $\h(\omega, i) \in \mathbb{C}^{M}$  \cite{brandstein2013microphone, benesty2008microphone} 
towards the $i^\text{th}$ candidate location relative to the first microphone is solely 
defined by  $\Delta t_{m,1}(i)$, $m = 2,\dots M$ as
\begin{align}
\h(\omega, i) &= \bigl[1 \,\, e^{-j\omega\Delta  t_{2,1}(i)}\,\, \dots\,\, e^{-j\omega\Delta t_{M,1}(i)}\bigr]^\transp. \label{h_steering}
\end{align}
With $\bPsi(\omega)$ in (\ref{corrmatrix}) and $\h(\omega, i)$ in (\ref{h_steering}), we can define the
 frequency-dependent SRP map
\begin{align}
\mathit{SRP}(\omega,i) &= \h^\herm(\omega, i)\bPsi(\omega)\h(\omega, i) - \tr[\bPsi(\omega)]\nonumber\\
%&\quad - \tr[\bPsi(\omega)]
&= 2 \hspace{-.45cm}\sum_{{\hspace{.2cm}\scriptscriptstyle (m,m'): m>m'\hspace{-.2cm}}}\hspace{-.45cm} \Re\bigl[\psi_{m,m'}(\omega)e^{j\omega\Delta t_{m,m'}(i)}\bigr] \label{SRP_omegaT}
\end{align}
where $\tr[\bPsi_{{y}}(\omega)]$ is commonly subtracted  \cite{dibiase2000high, velasco2016proposal, zhang2008does, zotkin2004accelerated, dmochowski2007generalized, do07, do2007fast, do09, cobos2010modified, marti13, nunes14} as it constitutes an offset independent of $i$ and hence does not contribute to the objective of spatial mapping.
The second step in (\ref{SRP_omegaT}) is obtained from the relation $\Delta t_{m,1}(i) - \Delta t_{m',1}(i) = \Delta t_{m,m'}(i)$ and the Hermitian nature of $\bPsi(\omega)$.
Here, the sum is to be taken over all combinations $(m,m')$ with $m>m'$, i.e. over all $P$ microphone pairs.
From (\ref{SRP_omegaT}), the broadband SRP value is obtained by integrating $\mathit{SRP}(\omega, i)$ over frequency, 
\begin{align}
\mathit{SRP}(i) &= \int_{-\omega_0}^{\omega_0}\hspace{-.15cm}\mathit{SRP}(\omega, i)\,d\omega. \label{SRP_T}
\end{align}
where the integration limits coincide with the assumed bandlimit. % of the microphone signals.
The location of a single source can be estimated by finding the candidate location index $i$ that maximizes $\mathit{SRP}(i) $.

\subsection{Relation to Time-Domain GCC}
\label{sec:timedomainGCC}
The integral in (\ref{SRP_T}) may be reformulated into a simple relation between $\mathit{SRP}(i)$ and the time-domain GCCs of the $P$ microphone pairs \cite{dibiase2000high, velasco2016proposal}.
Let $\xi_{m,m'}(\tau)$ denote the time-domain GCC defined by the IFT of the frequency-domain GCC $\psi_{m,m'}(\omega)$, i.e.
\begin{align}
\xi_{m,m'}(\tau) &= \int_{-\omega_0}^{\omega_0}\hspace{-.15cm}\psi_{m,m'}(\omega)e^{j\omega\tau}\,d\omega. %\nonumber \\
%&= \mathfrak{F}^\inv[\psi_{m,m'}(\omega)](\tau).
 \label{GCC_IFT}
\end{align}
Then, using (\ref{SRP_omegaT}) in (\ref{SRP_T}) and (\ref{GCC_IFT}), one finds that $\mathit{SRP}(i)$ becomes
\begin{align}
\mathit{SRP}(i) &= 2\hspace{-.45cm}\sum_{{\hspace{.2cm}\scriptscriptstyle (m,m'): m>m'\hspace{-.2cm}}}\hspace{-.45cm} \xi_{m,m'}(\Delta t_{m,m'}(i)),
 \label{SRP_T_IFT}
\end{align}
where one can verify that the real-part operator may be dropped by taking it out of the integral and acknowledging that $\psi_{{m,m'}}(\omega)$ is complex symmetric. 
Regarding (\ref{GCC_IFT})--(\ref{SRP_T_IFT}), recall that $\Delta t_{m,m'}(i) \in [-\Delta t_{m,m'|0} ,\, \Delta t_{m,m'|0}]$ according to (\ref{eq:Deltat_i_range}), while $\xi_{m,m'}(\tau)$ itself has unlimited support in $\tau$.

\subsection{Computational Complexity}
\label{sec:compcomp}
In conventional SRP, a discrete-frequency variant of the IFT in (\ref{GCC_IFT})
is evaluated for each $\tau = \Delta t_{m,m'}(i)$, with the integral replaced by a summation. %which is
%typically done by approximating the integral by a summation over discrete frequency bins.
Given  $\psi_{{m,m'}}(\omega)$ at $K$ frequency bins, this computation requires %the computation of (\ref{SRP_omegaT})--(\ref{SRP_T}) requires  
\begin{align}
C_{\textsl{conv}} = JPK \label{eq:SRPomp}
\end{align}
(complex) multiplications.
Both $J$ and $K$ are design parameters and commonly chosen in the order of hundreds or thousands, making conventional SRP rather complex. E.g., for a $D$-dimensional grid of candidate locations with 100 grid points per dimension, we obtain $J=100^D$.

\section{Complexity Reduction based on Nyquist-Shannon Sampling}
\label{sec:compred}
In order to reduce the computational complexity of SRP, we make use of the time-domain GCC relation in Sec. \ref{sec:timedomainGCC}.
In Sec. \ref{sec:samplinginterpolation}, we first restate the time-domain GCCs in terms of the Nyquist-Shannon sampling theorem, exploiting bandlimitedness. 
In Sec. \ref{sec:lcapprox}, we propose to approximate the time-domain GCCs and in turn the SRP map by interpolation from a limited the number of samples, thereby limiting the number of IFT computations and hence reducing computational complexity, as analyzed in Sec. \ref{sec:complApprox}. 

\subsection{GCC Sampling and Interpolation}
\label{sec:samplinginterpolation}

Recall that $\xi_{m,m'}(\tau)$ is bandlimited by $\omega_0$, as can be seen from (\ref{GCC_IFT}).
We can therefore invoke the Nyquist-Shannon sampling theorem \cite{marks2012introduction}.  
Let $T$ denote the critical sampling period \cite{marks2012introduction}, i.e.
\begin{align}
T = \pi/\omega_0,
\end{align}
and let $\xi_{m,m'}(\tau)$ be sampled at $\tau = nT$ with $n \in \mathbb{Z}$.
Using Whittaker-Shannon interpolation  \cite{marks2012introduction}, we can then express $\xi_{m,m'}(\tau)$ as a function of its samples $\xi_{m,m'}(nT)$, 
\begin{align}
\xi_{m,m'}(\tau) = \sum_{n = -\infty}^{\infty} \xi_{m,m'}(nT) \sinc(\tau/T - n), \label{eq:SRPinterp}
\end{align}
permitting perfect reconstruction. 

\subsection{GCC and SRP Approximation}
\label{sec:lcapprox}
As $\xi_{m,m'}(\tau)$ is bandlimited in $\omega$, it has unlimited support in $\tau$  \cite{marks2012introduction} and thus, as stated by (\ref{eq:SRPinterp}), one indeed needs infinitely many samples  $\xi_{m,m'}(nT)$ in order to exactly reconstruct $\xi_{m,m'}(\tau)$.
Nonetheless, we may approximate $\xi_{m,m'}(\tau)$ based on (\ref{eq:SRPinterp}) and the following arguments.

In the SRP use case, the approximation needs to be accurate only within the interval $\tau \in[-\Delta t_{m,m'|0} ,\, \Delta t_{m,m'|0}]$,  cf. (\ref{eq:Deltat_i})--(\ref{eq:Deltat_i_range}), where $\tau$ may be interpreted as the TDOA of a physical source location. 
Samples $\xi_{m,m'}(nT)$ sufficiently far {outside} this interval, i.e. at $\lvert n \rvert  \gg \Delta t_{m,m'|0}/T$, have little effect {within} the interval for two reasons. %on $f(\Delta t)$ within the interval for two reasons.
First, their effect is limited by nature of the interpolating sinc-function, whose envelope decays with the inverse of its argument.
Second, also the samples $\xi_{m,m'}(nT)$ themselves decay 
as the GCC lag $nT$ does not correspond to a TDOA anymore.\footnote{Exceptions due to, e.g., strong reflections or source-signal periodicities are possible, but irrelevant to the source localization problem.} % implies that the beampattern of $\h(\omega, nT)$ does not have a mainlobe, but only sidelobes. 
We therefore propose to approximate $\xi_{m,m'}(\tau)$ by truncating (\ref{eq:SRPinterp}) as
\begin{align}
\xi_{m,m'|\textsl{appr}}(\tau) &=\hspace{-1.5cm}\sum_{\hspace{1.5cm}n = -N_{m,m'} + N_{\textsl{aux}}}^{\hspace{1.3cm}N_{m,m'} + N_{\textsl{aux}}} \hspace{-1.5cm}\xi_{m,m'}(nT) \text{sinc}(\tau/T - n),\label{eq:SRPapprox}\\
\text{with}\,\,\,\,\, N_{m,m'} & = \left\lfloor \Delta t_{m,m'|0}/T \right\rfloor, 
\end{align}
where 
$ N_{\textsl{aux}}$ is a design parameter defining the number of auxiliary sample points beyond $[-\Delta t_{m,m'|0} ,\, \Delta t_{m,m'|0}]$. 
In practice, few auxiliary samples are sufficient, cf. also Sec.~\ref{sec:sim} and the examplary plots in Fig. \ref{fig:ex}, where $\xi_{m,m'}(\tau)$ and $\xi_{m,m'|\textsl{appr}}(\tau)$  are shown for $N_\textsl{aux} = 0$ and $N_\textsl{aux} = 2$. 

 \setlength\fwidth{7.3cm}
 \begin{figure}
\centering
\hspace*{-0.35cm}
    \setlength\fheight{5cm} 
    \input{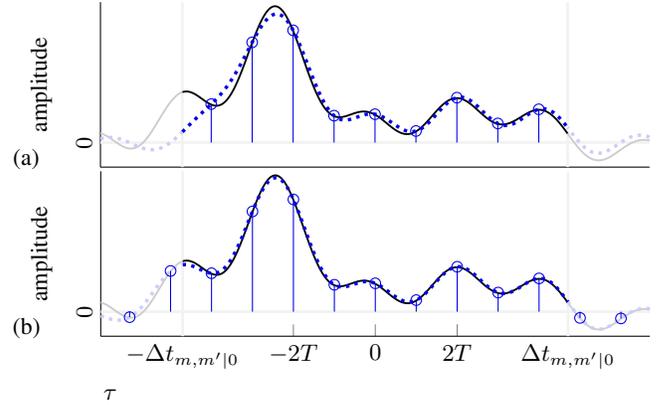} 
\caption{\footnotesize{Example of the GCC  $\xi_{m,m'}(\tau)$  [\ref{SRP_LC_ex_circ_1}] and its approximation $\xi_{m,m'|\textsl{appr}}(\tau)$  [\ref{SRP_LC_ex_circ_2}] with correlation lag $\tau  \in [-\Delta t_{m,m'|0},\, \Delta t_{m,m'|0}]$ for a microphone distance of $d_{m,m'} = 10\,\si{cm}$. 
Interpolation is performed from $\xi_{m,m'}(nT)$ [\ref{SRP_LC_ex_circ_5}] with (a) $N_\textsl{aux} = 0$ and (b) $N_\textsl{aux} = 2$ auxiliary samples. The microphone signals are generated as described in Sec. \ref{sec:sim}.} }
\label{fig:ex}
\end{figure}

Based on the GCC approximation discussed in Sec. \ref{sec:lcapprox}, the SRP approximation is straightforwardly obtained by  first evaluating $\xi_{m,m'}(\tau)$ 
at sample points $\tau = nT$ using a discrete-frequency variant of (\ref{GCC_IFT}), then obtaining $\xi_{m,m'|\textsl{appr}}(\tau)$ at $\tau=\Delta t_{m,m'}(i)$ using
 (\ref{eq:SRPapprox}), and summing  $\xi_{m,m'|\textsl{appr}}(\Delta t_{m,m'}(i))$ over all microphone pairs using (\ref{SRP_T_IFT}).
In the following, let this SRP approximation  be denoted by $\mathit{SRP}_\textsl{appr}(i)$.

\subsection{Computational Complexity}
\label{sec:complApprox}
According to (\ref{eq:SRPapprox}), the approximation $\xi_{m,m'|\textsl{appr}}(\tau)$ requires $2N_{m,m'} + 2N_\textsl{aux} + 1$ samples.
Then, per microphone pair, the computation of $\mathit{SRP}_\textsl{appr}(i)$ on average requires 
\begin{align}
N = \frac{ 2}{ P}\hspace{-.4cm}\sum_{{\hspace{.25cm}\scriptscriptstyle (m,m'): m>m'\hspace{-.25cm}}}\hspace{-.45cm}N_{m,m'} + 2N_\textsl{aux} + 1
\end{align}
samples.
Given  $\psi_{m,m'}(\omega)$ at $K$ frequency bins, the computation of all samples $\xi_{m,m'}(nT)$ for all $P$ microphone pairs using a discrete-frequency variant of the IFT in (\ref{GCC_IFT}) requires 
\begin{align}
C_{\textsl{samp}} = NPK \label{eq:SRPapproxcompC}
\end{align}
(complex) multiplications.
Given all $\xi_{m,m'}(nT)$, interpolating $\xi_{m,m'|\textsl{appr}}(\Delta t_{m,m'}(i))$ for all $P$ microphone pairs and all $J$ candidate locations using (\ref{eq:SRPapprox}) further requires  
\begin{align}
C_{\textsl{int}}  = NPJ  \label{eq:SRPapproxcompR}
\end{align}
(real) multiplications, where the weights $\sinc(\Delta t_{m,m'}(i)/T - n)$ can be pre-computed given the array geometry and the candidate location grid.
With (\ref{eq:SRPomp}) and (\ref{eq:SRPapproxcompC})--(\ref{eq:SRPapproxcompR}), we define (for simplicity, assuming that real and complex multiplications are equally expensive) the complexity reduction factor $R$ as
\begin{align}
R &= \dfrac{ C_{\textsl{samp}} + C_{\textsl{int}}}{ C_{\textsl{conv}}} =  R_{\textsl{samp}} + R_{\textsl{int}},\\
R_{\textsl{samp}} &= C_{\textsl{samp}}/C_{\textsl{conv}} = N/J,\\
R_{\textsl{int}} &= C_{\textsl{int}}/C_{\textsl{conv}} = N/K.
\end{align}
Here, $1 - R_{\textsl{samp}}$ indicates the saved cost in terms of IFT computations, while $R_{\textsl{int}}$ indicates the additional cost of interpolation.
In practical setups, a complexity reduction of several orders of magnitude may be obtained, as exemplified in the following.  
Assume that $c = 340\,\si{m/s}$ and $\nicefrac{\displaystyle \omega_0}{\displaystyle 2\pi} = 8\,\si{kHz}$, which is a commonly assumed bandlimit for speech signals, and consider a circular microphone array of $M=6$ equidistantly placed microphones. 
For this setup, Fig. \ref{fig:simRatio} shows $R_{\textsl{samp}}$ versus the array radius. % at different values of $N_{\textsl{aux}}$ and $J$. 
For a given $J$, we observe that $R_{\textsl{samp}}$ increases with the array radius. 
Note however that usually also $J$ scales with the array dimensions
(as for small dimensions, one would search the far field only, while for large dimensions, also the near field needs to be searched),
such that we would not find ourselves in the upper right region of the plot.
For $R_{\textsl{int}}$, a similar figure with $K$ in place of $J$ applies.\footnote{Note that instead of interpolating the SRP map for \textit{all} candidate locations, one could also employ refined spatial search approaches \cite{zotkin2004accelerated, do07, do2007fast, do09, cobos2010modified, marti13, nunes14}, yielding $R_{\textsl{int}} \ll N/K$. Such extensions are, however, beyond the scope of this paper.}

 \setlength\fwidth{7.9cm}
 \begin{figure}
\centering
\hspace*{-0.3cm}
    \setlength\fheight{4.2cm} 
    \input{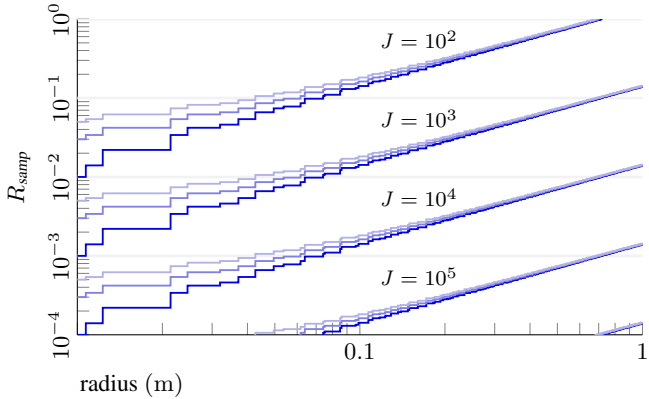} 
\caption{ \footnotesize{complexity reduction factor $R_{\textsl{samp}}$ for $N_{\textsl{aux}} = 0$ [\ref{SRP_Ratio_1}], $1$ [\ref{SRP_Ratio_2}], $2$ [\ref{SRP_Ratio_3}] auxiliary samples versus the radius of a circular microphone array of $M = 6$ equidistantly spaced microphones for different values of $J$.}}
\label{fig:simRatio}
\end{figure}

\section{Simulations}
\label{sec:sim}

In this section, we investigate the deviation of $\mathit{SRP}_{\textsl{appr}}(i)$ from $\mathit{SRP}(i)$ as well as the impact thereof on the localization error for different values of $N_\textsl{aux}$.
The acoustic scenario is modelled using the randomized image method \cite{desena15} at a sampling rate of $16\,\si{kHz}$. 
We simulate a shoe-box room of $6 \times 7 \times 3.5\, \si{m}$ with $600\,\si{ms}$ reverberation time and $0.89\,\si{m}$ critical distance, wherein a circular microphone array with $10\,\si{cm}$ radius  and $M = 6$ microphones, cf. also Sec. \ref{sec:complApprox}, as well as a single source are placed.
The microphone array lies in the horizontal plane $20\,\si{cm}$ below the ceiling, centered around $(2.9 \,\,\,3.4\,\,\,3.3)^\transp\,\si{m}$. 
Results are averaged for $256$ randomly selected source positions in at least $1\,\si{m}$ distance to microphone array.
Both female and male speech \cite{bando92} are used as source signals. 
Diffuse babble noise \cite{habets2008generating, Audiotec} is added at signal-to-noise ratios (SNRs) of $\mathit{SNR} = \{-3,\,0,\,3, \,6\}\,\si{dB}$. 
Per source location and SNR, we generate $2.11\,\si{s}$ of audio data. 
For the given microphone array, with $c = 340\,\si{m/s}$ and $\nicefrac{\displaystyle \omega_0}{\displaystyle 2\pi} = 8\,\si{kHz}$, we obtain $P = 15$ and  $N =  14.2 + 2N_{\textsl{aux}}$. 
Assuming far field propagation, a spherical grid of candidate locations is employed, covering the lower half-sphere.
With  $\theta_i$ and $\phi_i$ denoting the polar and azimuthal candidate incident angles with zenith and azimuth direction along the third (i.e., the vertical) and the first room dimension, respectively, such that $ \mathbf{r}_{i} =  -(\sin \theta_i \cos \phi_i  \,\,\, \sin \theta_i \sin \phi_i  \,\,\, \cos \theta_i )^\transp$, we vary $\theta_i$ between $90^\circ$ and $180^\circ$ with $2^\circ$ resolution, and $\phi_i$ between $0^\circ$ and $358^\circ$ with $2^\circ$ resolution.
In total, this yields $J = 8101$ different candidate locations.
The frequency-domain GCC in (\ref{corrmatrix}) is implemented in frames by means of the short-time Fourier transform (STFT) using square-root Hann windows of $2048$ samples, resulting in $K = 1024$ (considering one side of the spectrum only).
In terms of computational complexity, we hence obtain $C_{\textsl{conv}} \approx 124.43 \cdot 10^6$ for conventional SRP, and, choosing $N_\textsl{aux} = 2$ for instance, $C_{\textsl{samp}} \approx 0.28 \cdot 10^6$ and $C_{\textsl{int}}  \approx 2.21 \cdot 10^6$ for the proposed approximation, yielding $R_{\textsl{samp}} \approx 2.25 \cdot 10^{-3}$ and $R_{\textsl{int}} \approx 1.78 \cdot 10^{-2}$. 
We define the approximation error of $\mathit{SRP}_{\textsl{appr}}(i)$ w.r.t. $\mathit{SRP}(i)$ by 
\begin{equation}
e_\textsl{appr} = 10\log_{10} \frac{ \sum_i \bigl( \mathit{SRP}(i) - \mathit{SRP}_{\textsl{appr}}(i)\bigr)^2}{ \sum_i \mathit{SRP}^2(i)}\,\si{dB},
\end{equation}
and define the localization error $e_\textsl{local}$ as the angle between the true and the estimated incident direction. 

Fig. \ref{fig:sim} reports the simulation results in terms of median errors (the median is preferred over the mean since the distribution of $e_\textsl{local}$ is heavy-tailed). 
In Fig. \ref{fig:sim} (a), the median of $e_\textsl{appr}$  [\ref{SRP_LC_V5_5}] over all source locations, all SNRs, and all frames versus $N_\textsl{aux}$ is shown, with shaded areas indicating the range from the first to the third quartile.
Fairly good approximations with $e_\textsl{appr} \approx -31.5\,\si{dB}$ are obtained at $N_\textsl{aux} = 0$ already; the error reduces further to $e_\textsl{appr} \approx -34\,\si{dB}$ at $N_\textsl{aux} = 2$.
Fig. \ref{fig:sim} (b) depicts the median  localization error $e_\textsl{local}$ over all source locations and all frames using $\mathit{SRP}(i)$ [\ref{SRP_LC_V5_7}] and $\mathit{SRP}_{\textsl{appr}}(i)$ [\ref{SRP_LC_V5_10}]  versus $N_\textsl{aux}$ at various SNRs. %for comparison.
Naturally, better estimates are obtained for higher SNR values.
While a performance difference of $0.8^\circ$ to $1.8^\circ$ between $\mathit{SRP}(i)$  and  $\mathit{SRP}_{\textsl{appr}}(i)$ can be observed at  $N_\textsl{aux} = 0$, the additional accuracy loss due to the approximation drops to $0^\circ$ at $N_\textsl{aux} = 2$ already. 
Obviously, even for $N_\textsl{aux} = 0$, the SNR has a much larger impact on the localization performance than the approximation error.

 \setlength\fwidth{7.2cm}
 \begin{figure}
\centering
\hspace*{-0.3cm}
    \setlength\fheight{6.1cm} 
    \input{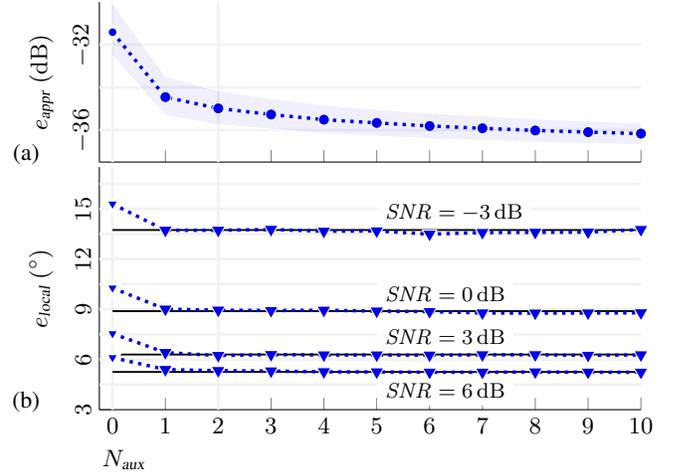} 
\caption{\footnotesize{(a) Median approximation error $e_\textsl{approx}/\si{dB}$  [\ref{SRP_LC_V5_5}] versus the number of auxiliary samples $N_\textsl{aux}$; (b) median localization error $e_\textsl{local}/^\circ$ using $\mathit{SRP}(i)$ [\ref{SRP_LC_V5_7}] and $\mathit{SRP}_{\textsl{appr}}(i)$ [\ref{SRP_LC_V5_10}] versus the number of auxiliary samples $N_\textsl{aux}$ at various SNRs.}}
\label{fig:sim}
\end{figure}

\section{Conclusion}
\label{sec:conclusion}

As an alternative to computationally complex conventional SRP, we have proposed a low-complexity approximation based on the Nyquist-Shannon sampling.
Commonly, the number of required samples is orders of magnitude less than the number of candidate locations and frequency bins, allowing for significant complexity reduction at an equal localization performance.

\bibliographystyle{IEEEtran}
\bibliography{/Users/thomasdietzen/Syncbox/DOC/bib/MyResearch}

\end{sloppy}
\end{document}